# High-directivity multi-level beam switching with single-gate tunable metasurfaces based on graphene

Juho Park[1], Ju Young Kim[1], Sunghyun Nam[1], and Min Seok Jang[1*]

[1]School of Electrical Engineering, Korea Advanced Institute of Science and Technology, Daejeon 34141, Republic of Korea

*Corresponding author. Min Seok Jang: jang.minseok@kaist.ac.kr

**Abstract**

 The growing demand for ultra-fast telecommunications, autonomous driving, and futuristic technologies highlights the crucial role of active beam steering at the nanoscale. This is essential for applications like LiDAR, beam-forming, and holographic displays, especially as devices reduce in form-factor. Although device with active beam switching capability is a potential candidate for realizing those applications, there have been only a few works to realize beam switching in reconfigurable metasurfaces with active tuning materials. In this paper, we theoretically present a multi-level beam-switching dielectric metasurface with a graphene layer for active tuning, addressing challenges associated with achieving high directivity and diffraction efficiency, and doing so while using a single-gate setup. For two-level switching, the directivities reached above 95%, and the diffraction efficiencies were near 50% at the operation wavelength $\lambda_0$ = 8 μm. Through quasi-normal mode expansion, we illustrate the physics of the beam switching metasurface inverse-designed by the adjoint method, highlighting the role of resonant modes and their response to charge carrier tuning. Under the same design scheme, we design and report characteristics of a three-level and four-level beam switching device, suggesting a possibility of generalizing to multi-level beam switching.

**Keywords**: Beam switching, active metasurfaces, graphene, mid-infrared, photonic inverse design, adjoint method, free-form optimization

# 1. Introduction

As the demand for ultra-fast telecommunications, autonomous driving, and other futuristic technologies continues to grow, the need for wavefront shaping capability at the micro-scale has become more evident [1–4]. Metasurfaces, consisting of subwavelength dielectric or plasmonic components, are suitable candidates for realizing complex light manipulations. To date, many metasurface designs have been static, i.e. the geometric and optical parameters of the underlying components are fixed after fabrication. However, advanced optical applications, such as LiDAR [5], tunable metalenses [6], wavefront shapers [7], and holographic displays for virtual/augmented reality [8], need active/reconfigurable metasurfaces that can be tuned dynamically. Active beam-switching is one such application.

To enable such active beam-switching, various forms of reconfigurable metasurfaces have been suggested [9], with the tuning mechanisms spanning electro-optic, thermo-optic, and mechanical modulations. Solid state methods, such as charge carrier tuning, is optimal for the reasons of local controllability and robustness, while on the other hand using a dielectric meta-atom platform is preferred over the usage of metallic components, which has significant ohmic dissipations. These conditions naturally pinpoint to the usage of a dielectric platform with a charge carrier-tunable 2D material, which has a minimal volume that is lossy, yet is required for the active tuning. Among these, graphene charge carrier tuning is a favorable candidate for active tuning in reconfigurable metasurfaces, owing to graphene's atomic layer thickness and high electrostatic tunability [10]. There have been various demonstrations of active metasurfaces highlighting graphene's promising tuning capability, such as active thermal emission control [11,12] and high-efficiency independent modulation of phase and amplitude [13,14]. The tuning speed, or the switching rate, of the GHz range in graphene [15] is also superior to that of the other methods such as the usage of liquid crystals, which only has a switching rate in the KHz range [16].

Given the active tuning method, there are various ways of designing the metasurface for manipulating the wavefront into desired diffraction orders. Generally speaking, these can be categorized into bottom-up methods and top-down methods. Bottom-up methods include the unit-cell method, which spatially arranges designated meta-atoms according to the required phase profile [17]. For the designing of active metasurfaces, unit-cell method often requires a multi-gated setup for dynamically tuning each meta-atom separately according to the desired beam deflecting angle, leading to a complex experimental arrangement [14,18]. Also, it is widely known that metasurfaces designed by the unit-cell method likely yield limited efficiencies due to the anticipated mode profiles of the unit cells overlapping with the neighboring ones, leading to significant deviation from the locally periodic approximation and resulting in undesired wavefronts [19,20]. This phenomenon is aggravated in cases requiring abrupt metasurface field manipulations, such as implementing high numerical aperture in metalenses or large beam diffraction angles in metagratings [21]. Nevertheless, due to its simple design procedure and modeling capabilities, it is prominently used in many metasurface design works [22-25]. On the opposing end, the top-down, holistic design methods use nonlocalized modes to enable high-efficiency beam diffraction under abrupt field manipulation [26], incorporating the overlapping field effects between neighboring meta-atom components. The top-down methods include the adjoint-based topology optimization [27] or deep-learning based method [28]. There are numerous publications regarding high-performance beam deflecting metasurfaces designed by these methods [29-33].

Among the numerous qualities of beam switching metasurfaces, we focus on satisfying the key performance metrics simultaneously, namely the high beam diffraction angle, high directivity, and diffraction efficiency for multi-level switching. Designing a dynamic high-angle beam switching metasurface compounds the complexities, and we overcome the associated difficulties by using holistic methods. We observed that the current state-of-the-art active beam switching metasurfaces require a multi-gated setup or have low beam diffraction angles. In this paper, we aim to design a metasurface that improves those aspects, so that the needs of achieving high directivities/efficiencies for high beam switching angles, using a single-gated setup, and implementing multi-level beam switching are all met. Our metagrating setup is a dielectric one, with a graphene layer for active tuning purposes, which minimizes energy loss while keeping the freedom to significantly tune charge carriers. Utilizing the adjoint-based holistic optimization method, we first demonstrate a two-level beam switching metasurface, for the sake of simply illustrating the physics of our metagrating, which involves the utilization of different quasinormal modes (QNMs) and how they respond to the charge carrier tuning. The pole-zero analysis and QNM expansion method are used to trace the trajectories of these resonances through the corresponding poles and elucidate their roles in beam switching [34]. Finally, as a trial toward generalized multi-level beam switching, we move on to the design of a three- and four- level beam switching device, where the principles behind how the different QNMs enable the beam switching are similar but involves more QNMs.

## 2. Results and discussion

### 2.1 Metasurface configuration

In this section, we propose a reflective active beam switching metasurface utilizing an unpatterned graphene as a tuning material. **Fig. 1** illustrates the unit-cell structure of the three-level switching metasurface and beam switching operation. The structure consists of a Si metagrating with thickness $t_g$, an unpatterned graphene layer, a silicon nitride spacer with thickness $t_{spacer}$, a Si substrate with thickness $t_s$, and an Au back reflector. The period is set to $\Lambda=\lambda_0/\sin(\theta)$ in order to eliminate higher diffraction orders and force the +1$^{st}$ and -1$^{st}$ diffraction orders to have diffraction angles of $\pm\theta$, where $\theta = 80°$ throughout this paper. The incident plane wave is transverse-electric (TE) polarized to minimize the ohmic losses in the back reflector. The operating wavelength is set to $\lambda_0 = 8$ μm, which causes both carrier intraband and interband transition in graphene within our Fermi energy range of interest [35]. The Si metagrating, the principal region to be optimized, acts as a coupling interface, connecting the normally incident plane waves into desired diffraction orders. Acting as a gate dielectric for preventing electrical shorts, the thin silicon nitride spacer ($t_{spacer} = 30$ nm) is inserted between the graphene layer and Si substrate. The Au back reflector enhances the light-matter interaction by doubling the light path. The graphene and the Au back reflector are connected to apply gate biases ($V_g$). Upon different applied gate voltages, the three-level beam switching metasurface reflects the incident plane wave into either +1$^{st}$ (**Fig. 1a**), 0$^{th}$ (**Fig. 1b**), or -1$^{st}$ (**Fig. 1c**) diffraction channels.

### 2.2 Inverse design of two-level beam switching metasurface

To gain an insight into how active beam switching works, we first demonstrate a two-level beam switching metasurface. For the two-level beam switching metasurface, unit-cell period is set to $\Lambda$, and the Fermi energy switching levels are 0.2 and 0.6 eV, respectively. To achieve a clear plane wavefront with high deflection, we utilize the adjoint method as a design strategy.

Along with classical optimization methods [36,37] and deep-learning based design methods [38,39], the adjoint method is prevalently used for many photonic inverse design problems [31,40,41], owing to its efficient evaluation of the FoM (figure of merit) gradient upon design parameters. Specifically, it enables the calculation of the FoM gradient using only two simulations, forward and adjoint simulations, regardless of the number of design parameters [42,43]. A more detailed explanation on the adjoint gradient formulation is given in **Supplementary Information Section S1**. There are two types of adjoint gradients. One is the grayscale gradient which is the gradient when the permittivity value changes while the shape is fixed. The other is the shape-derivative which is the gradient when the shape changes while the permittivity is fixed. Deploying the shape-derivative instead of the grayscale adjoint gradient [40] enables one to naturally control the highest aspect ratio, by integrating it as a constraint to the optimization problem. In this work, we set the highest aspect ratio of Si pillars to be 10, which is realizable under the use of contemporary etching techniques used in mass-production, including deep reactive ion etching [44].

For beam switching/steering applications, diffraction efficiency is an important characteristic as it is directly related to the energy efficiency of the device. Also, beam directivity is a crucial factor as it is directly related to beam divergence and spatial resolution [18,45]. Therefore, it is important for a beam switching metasurface to maximize both performance metrics simultaneously. Throughout this paper, we denote the diffraction efficiency into $i^{\text{th}}$ diffraction channel as $R_i$, as there is no transmission through the metasurface. Directivity is defined as the ratio of the diffraction efficiency into the target diffraction channel ($R_{\text{target}}$) divided by sum of diffraction efficiencies: $D_{\text{target}} = R_{\text{target}} / \sum_i R_i$. To satisfy high directivities and high diffraction efficiencies for all switching states, we define the total FoM to be maximized as follows:

$$\text{FoM} = \sum_{\text{states}} (b R_{\text{target}} - \sum_{i=\text{off-target}} R_i),$$

where $b$ is a positive constant. Setting $b$ to be small relatively puts less weight on the target diffraction efficiency and more weight on the off-target efficiencies. In this case, the optimizer emphasizes minimizing the off-target diffraction efficiencies (due to the minus sign in the FoM) at the expense of maximizing the target diffraction efficiency, leading to a high directivity as the off-target diffraction efficiencies are greatly minimized but with the cost of a slightly lower target diffraction efficiency for each switching state. On the other hand, a higher $b$ value relatively puts heavier weight on the target diffraction efficiency and less on the off-target efficiencies. This results in a high target diffraction efficiency but low directivity. Balancing this trade-off, we subtly tune $b$ to 0.3 in order to achieve both high directivity and diffraction efficiencies.

Using the aforementioned optimization procedures, we sweep the thicknesses of the metagrating layer $t_g$ and substrate layer $t_s$ from $k_{\text{Si}} t_g = \pi/8$ to $2\pi$ and $k_{\text{Si}} t_s = \pi/8$ to $2\pi$ with steps of π/8, where $k_{Si} = n_{Si}k_0$ and $k_0$ is the free-space wavevector at $λ_0$ = 8 μm. For each pair of ($t_g$, $t_s$), five optimizations were conducted with different initial points, leading to a total of 1280(=256×5) optimizations. By examining the sweep results, we found that there exist "forbidden zones" which prevent achieving both high diffraction efficiencies and directivities, comprehensively analyzed in **Supplementary Information Section S2**. Interestingly, for two-, three-, and four-level beam switching device optimizations, the designs with both high directivities and high diffraction efficiencies were often located near the point of ($k_{Si}t_g$, $k_{Si}t_s$) = (3π/2, 3π/2). Therefore, throughout this paper, we report the case of $t_g$ = $t_s$ = 3π/(2$k_{Si}$) = 1.75 μm.

**Figure 2** depicts the geometrical structure of the optimized metasurface and the optical responses. The unit cell structure of the optimized two-level beam switching metasurface is shown in **Fig. 2a**. The exact geometric parameters of optimized structures for two-, three-, and four-level switching metasurfaces are shown in **Supplementary Information Section S5**. At $E_F$ = 0.2 eV the optimized metasurface shows directivity and DE of 96.6% and 52.4%, respectively, and at $E_F$ = 0.6 eV, 97.0 % and 49.4%, respectively. The diffraction efficiencies along different graphene Fermi levels at the operating wavelength $λ_0$ = 8 μm are shown in **Fig. 2b**, whereas **Fig. 2c** shows the DEs for different wavelengths at two switching states. The scattered electric field simulated by the FEM solver reveals clear wavefronts towards the target directions (**Fig. 2d**). The angles of the formed wavefronts manifest in ±80° which correspond to desired diffraction angles.

Here, we focus on the observation that the dependence of reflectance on the Fermi levels are like the dependence on the wavelengths (**Fig. 2b-c**). The similarity can be explained by investigating the resonance frequency tuning mechanism. From perturbation theory [46], the resonance frequency shift Δω can be approximated to the first-order, leading to:

$$\Delta\omega = -\frac{\omega_0}{2}\frac{\int dv\, \Delta\epsilon(r)|E(r)|^2}{dv\, \epsilon(r)|E(r)|^2} \quad (1)$$

where $ω_0$ and $E(r)$ are the mode resonance frequency and its electric field, respectively, and $\epsilon(r)$ and $\Delta\epsilon(r)$ are the permittivity distribution and its change due to active control, respectively. The integral operator is applied to the whole region of the unit-cell. In our case, the active control parameter corresponds to the Fermi level of graphene. As the Fermi level changes, changes in permittivity occur only in the active material region, i.e. unpatterned graphene layer. Also, the electric fields are dispersed over a larger area of the whole metasurface, unlike the case of graphene plasmons, which involve extremely highly concentrated electric fields around the graphene regions [13]. Therefore, the field intensities are "weak" enough for the first-order perturbation approximation **Eq. 1** to uphold, which results in the resonance frequency shift $\Delta\omega$ being roughly proportional to the changes in permittivity $\Delta\epsilon$. Despite the nonlinearity in the dependence of graphene $\Delta\epsilon$ on the Fermi level change $\Delta E_F$ [47], the actual plot of the complex $\Delta\epsilon$ on $\Delta E_F$ yields a linear-looking curve at $\lambda_0 = 8$ μm, room temperature, and mobility values of 500-2000 cm$^2$/V·s (see **Supplementary Information Section S6**). As a result, the resonance frequency shift $\Delta\omega$ shows almost linearly proportional behavior to the Fermi level change $\Delta E_F$, i.e. $\Delta\omega \propto \Delta\epsilon \propto \Delta E_F$. This linear resonance frequency shift with respect to the graphene Fermi level has also been observed in quasi-bound states in the continuum (qBIC) tuning in unpatterned graphene based metasurfaces [13]. Since this linear resonance frequency shift occurs continuously while preserving the resonance lineshapes, the diffraction efficiency spectrum drawn as a function of $\lambda$ (**Fig. 2c**) consequently resembles the one drawn as a function of $E_F$ (**Fig. 2b**). Also, for the same reason, the diffraction efficiency spectrum (**Fig. 2c**) for $E_F = 0.6$ eV is similar to the one for $E_F = 0.2$ eV, with the zeros of diffraction efficiencies for $+1^{st}$ and $-1^{st}$ diffraction channels have been moved from 8 μm and $\lambda_1 = 8.0135$ μm ($E_F = 0.2$ eV) to $\lambda_2 = 7.989$ μm and 8 μm ($E_F = 0.6$ eV), respectively.

## 2.3 Quasinormal mode expansion of reflection coefficients

In this section, we analyze the reflection coefficients of each diffraction order in the two-level beam switching metasurface, by expanding them as a sum of the contributions of intrinsic QNMs. A QNM refers to an eigenmode when the system is open and non-Hermitian, in the presence of incoming and outgoing radiation channels and lossy materials in the system [48]. QNM mode expansion technique is a popular tool for investigating nanophotonic devices [49-52], owing to its capability of clearly identifying each QNM's contribution to certain optical responses. To illuminate how each of the QNMs, alongside the continuous radiation spectrum that they are coupled to, contributes to the spectrum of interest, we utilize Riesz Projection (RP) methods. RP methods have several advantages compared to other QNM expansion methods. First, RP methods do not require prior knowledge of the exact shape of eigenmodes, only the knowledge of eigenfrequencies. Second, they do not depend on orthogonality relations and are free from normalization issues. Third, they are based on contour integration, so the contributions of each mode can be related by a simple summation [51,53]. A brief formulation of the RP method is addressed in **Supplementary Information Section S7**.

As the RP method need evaluations of optical responses in the complex frequency plane, we used an open-source rigorous coupled-wave analysis (RCWA) solver S4 [54]. The material properties used for Si (substrate), $Si_3N_4$ (spacer), Au (back reflector) have been analytically extended to the complex plane by inserting complex frequency into their Drude-Lorentz model permittivity formula [55]. The conductivity of graphene has been analytically extended by using the conductivity model [56], which represents the conductivity as a function of frequency, temperature, and carrier density.

Before applying the QNM expansion, we first analyze the locations of the poles and zeros of $-1^{st}$, $0^{th}$, and $+1^{st}$ diffraction coefficients, denoted as $r_{-1}$, $r_0$, and $r_{+1}$, respectively, in the complex frequency plane as a function of Fermi levels. **Fig. 3a** illustrates the trajectories of the poles and zeros in the two-level beam switching metasurface. These poles and zeros locations are first estimated by a brute-force sweep in the complex frequency plane, and fine-tuned by local root-finding algorithms [57,58]. The locations of bands along the incident angle (photonic band structure) are reported in **Supplementary Information Section S3**. The pole locations are denoted by $\omega_{pn}$, where $n$ is the index of the poles numbered from low Re{$\omega$}, and indicated by black 'X' marks. The zero locations are denoted by $\omega_{zn,m}$, where $n$ is the index of the zeros numbered from low Re{$\omega$}, and $m$ represents the corresponding diffraction order. As the graphene Fermi level increases, all the poles and zeros tend to blueshift except $\omega_{z2,0}$, which almost remains unchanged. While the linear shift of poles along increasing Fermi levels is expected behavior, predicted by **Eq. (1)**, the linear shift of zeros is the result stemmed from complicated contributions of QNMs.

To achieve high directivities, all unwanted diffraction channels must be suppressed simultaneously by having zeros near the real target frequency $\omega_0 = 2\pi c/\lambda_0$, whereas all zeros of the target diffraction coefficient should be located far away from $\omega_0$. For our optimized device, as shown in **Fig. 3a**, $\omega_{z2,0}$ and $\omega_{z2,-1}$ are the two zeros located near $\omega_0$ and very close to the real frequency axis at $E_F = 0.2$ eV, making the majority of the beam diffracted into the $+1^{st}$ order. Similarly, at $E_F = 0.6$ eV, $\omega_{z2,0}$ and $\omega_{z1,+1}$ are near $\omega_0$, funneling incident light into the $-1^{st}$ order diffraction channel.

The complex pole-zero analysis also reveals that there exist two poles mainly involved in our frequency range of interest. The electric field profiles of the two QNMs corresponding to the two poles are drawn in **Fig. 3b**.

The QNM expansion illustrates how the two-level beam switching is achieved with the dynamics between these two QNMs and the background mode as displayed in **Fig. 3c**. The physical observables that are decomposed into the algebraic sum of the modal contributions are the diffraction coefficients, which are complex values, so we show both their amplitude and the phase. The summations obtained by the QNM expansion show good agreements with the reference reflection coefficients obtained by the RCWA solver. Slight differences between the summations and the reference values are possibly due to several reasons, such as inaccurate pole locations, insufficient number of divided segments in performing numerical contour integrals in the complex frequency space, or numerical error arising from different choices of contours containing the poles of interest.

As discussed above, to achieve high directivity beam switching, the +1st and -1st order diffraction channels must be alternatively switched on and off depending on $E_F$, whereas the 0th order channel must be nulled for both $E_F$ = 0.2 eV and 0.6 eV. For the +1st order diffraction channel, the net response of the resonant modes is dominated by the mode 1 as it mainly diffracts light into the +1st channel as shown in **Fig. 3b** and **3c**. At $E_F$ = 0.6 eV, the $r_{+1}$ amplitude of the mode 1 peaks around $\omega = \omega_0$ and its phase is ~π relative to the background diffraction. Similarly, even though it has a smaller amplitude, the mode 2 contribution is also nearly out-of-phase relative to the background diffraction. Since the combined amplitude of the mode 1 and mode 2 is similar to the background diffraction amplitude, the destructive interference between the background and the combined mode response is nearly complete, nullifying the +1st diffraction channel at $E_F$ = 0.6 eV. As the Fermi level changes to 0.2 eV, the relative phases of modes 1 and 2 with the background modes shift away from π, which veers away from destructive interference and results in a significant non-zero sum of ~0.7 in amplitude. Through similar destructive interference mechanism, -1st diffraction channel is suppressed at $E_F$ = 0.2 eV.

For the 0th diffraction channel, at $E_F$ = 0.2 eV, the background contribution is marginal and the resonant modes have a relative phase difference of roughly π with one another, explaining the destructive interference. At $E_F$ = 0.6 eV, it's more complex, as all three components (two resonant modes and the background mode) have comparable amplitudes, but their relative phase differences ultimately result in destructive interference.

## 2.4 Inverse design of three- and four-level beam switching metasurfaces

As an extension to the two-level beam switching metasurface, we design metagrating patterns to achieve three-level and four-level beam switching capabilities. For the three-level switching, the period is $\Lambda$, the same as in the two-level case. The Fermi levels of the switching states are set to $E_F = 0.2$, 0.6, and 1.0 eV for the +1st, 0th, and -1st diffraction orders, respectively. The optimized metagrating pattern is shown above in **Fig. 4a-b**, exhibiting 97.2%, 93.1%, and 98.0% directivities and 45.0%, 41.1%, and 40.5% diffraction efficiencies for $E_F = 0.2$, 0.6, and 1.0 eV respectively, as shown in **Fig. 4a**. The QNM expansion results for the three-level switching metasurface are discussed in **Supplementary Information Section S4**. The main difference between the two-level and the three-level beam switching metasurface is that the three-level beam switching one has three poles (QNMs). Also, the trajectories of the poles and zeros are more complicated than that of the two-level design, shown in **Fig. 4b**. Here, the mechanism of alternating zeros and nonzeros along the switching levels is similar to the two-level switching metasurface. At $E_F = 0.2$ eV, $\omega_{z3,0}$ and $\omega_{z3,-1}$ crosses $\omega = \omega_0$, making $R_0$ and $R_{-1}$ zero. At $E_F = 0.6$ eV, $\omega_{z3,+1}$ and $\omega_{z2,-1}$ crosses $\omega=\omega_0$, making $R_{+1}$ and $R_{-1}$ zero. At $E_F = 1.0$ eV, $\omega_{z1,+1}$ and $\omega_{z2,0}$ crosses $\omega=\omega_0$, making $R_{+1}$ and $R_0$ zero.

For the four-level switching, the period is set to $2\Lambda$, to have the extra -2nd and 2nd order diffraction channels. The Fermi levels for switching states are set to $E_F = 0.05$, 0.35, 0.65 and 0.95 eV for +2nd, +1st, -1st, and -2nd diffraction orders, respectively. The optimized metagrating pattern is displayed above **Fig. 4c-d** and it shows near 65% directivities for all the four states and 16.5%, 15.0%, 23.5%, and 26.0% diffraction efficiencies for $E_F = 0.05$, 0.35, 0.65 and 0.95 eV, respectively, as shown in **Fig. 4c**. In this case, the dynamic of the QNMs is too complex to unravel intuitively as the jumbled trajectories of many zeros and poles in **Fig. 4d** implies. From relatively low directivities compared to the two previous metasurface designs, we suspect that there might be a fundamental limitation for the possible number of switching states, and it was difficult for us to obtain higher performances for higher number of switching states as well.

Considering that the number of design objectives grows with the number of switching states and the number of diffraction orders, it naturally leads to the fact that it becomes increasingly more difficult to achieve higher multi-level beam switching. However, the findings reported in this analysis can be considered a preliminary result, and we suspect there might be ways to alleviate such complexities. For example, performing a larger-scale metasurface design may enable higher multi-level switching, by utilizing a larger design space and additional QNMs that could be used in sophisticated zero shapings.

## 3. Conclusion

In this work, we theoretically demonstrate single-gate, electrically tunable, multi-level beam switching metasurfaces, ranging from two to four switching levels. With the help of adjoint-based optimization, we successfully designed graphene-based active metasurfaces having both high directivities and diffraction efficiencies. For the two-level switching metasurface, the directivities reached above 95%, and the diffraction efficiencies were near 50% at the operation wavelength $\lambda_0 = 8$ μm. The operating mechanism is explained as the interference between the resonant modes and the non-resonant background modes through QNM expansion. Applying the same optimization method to the three-level switching metasurface led to above-90% directivities for all switching states. For the four-level switching metasurface design, the decline in the performance metrics raised the possibility of a fundamental trade-off/limitation on the number of switching states and the respective performances, which might be overcome by including more QNMs. Although the results presented in this paper are based on the mid-infrared frequency regime with graphene as a tuning material, the design framework and the analysis methods can be applied to other frequency regimes and general beam switching platforms. We hope this work provides to be a useful guide in high-performance tunable beam switching metasurface designs.

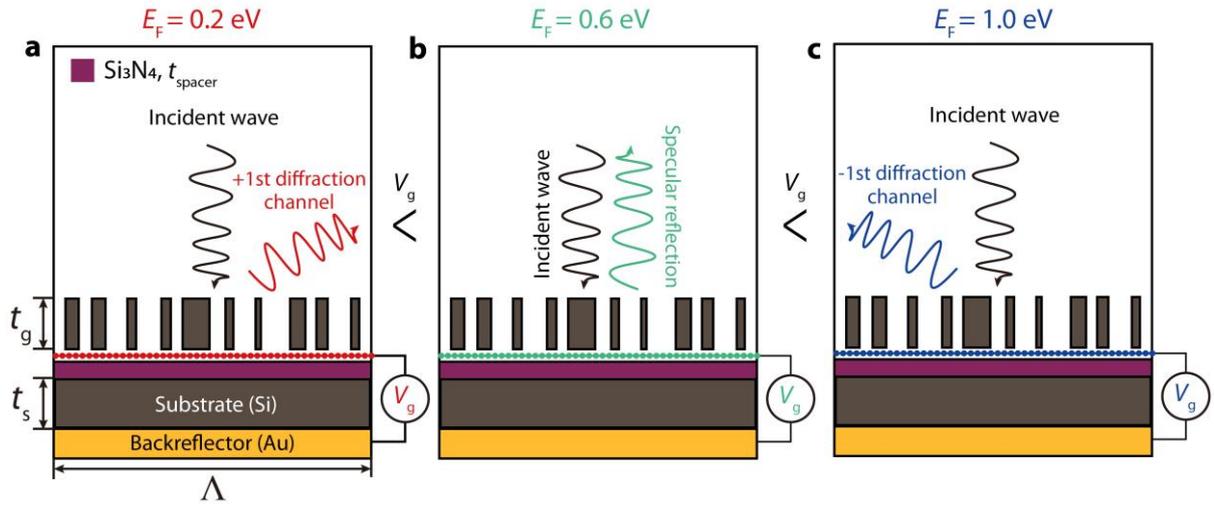

**Fig 1**. **Schematic of the three-level beam switching metasurface and the switching mechanism.** The free-form metagrating is located above the unpatterned graphene layer, serving as a resonant coupler, guiding normally incident wave into different diffraction channels. The gate voltage is applied between the Au back reflector and the unpatterned graphene. As the gate voltage increases, the metasurface switches the reflection of normally incident plane wave into **(a)** +1$^{st}$ diffraction order, **(b)** 0$^{th}$ diffraction order, and **(c)** -1$^{st}$ diffraction order. There is a thin Si$_3$N$_4$ spacer between the Si substrate and the graphene layer, as a gate dielectric.

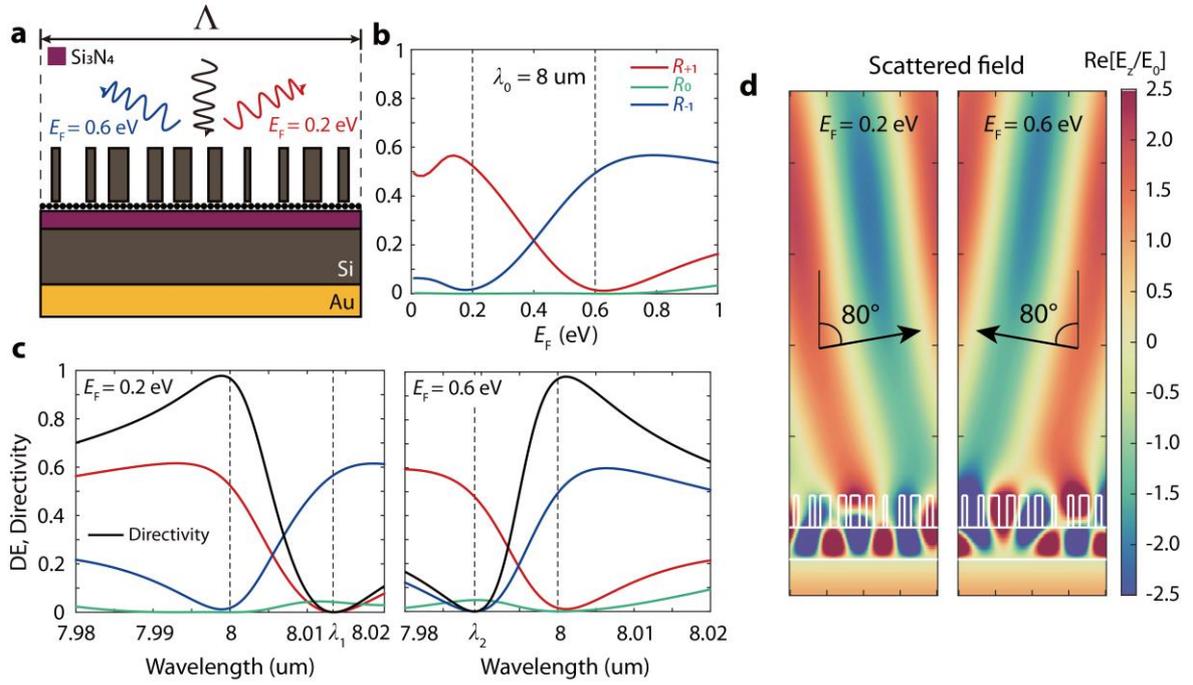

**Fig 2. Optimization result of two-level beam switching device. (a)** Unit-cell structure of the optimized two-level beam switching metasurface. Sizes are not to scale. **(b)** DEs of +1st (red solid line), 0th (green solid line), and -1st (blue solid line) diffraction orders at $\lambda_0$ = 8 μm, along graphene Fermi levels. The vertical dotted lines indicate the two switching Fermi levels, $E_F$ = 0.2 eV, and 0.6 eV. **(c)** DEs of +1st (red solid line), 0th (green solid line), and -1st (blue solid line) diffraction orders, and directivity (black solid line) for Fermi levels $E_F$ = 0.2 eV (left) and 0.6 eV (right), along wavelengths. The vertical dotted lines indicate the zeros of DEs for +1st and -1st diffraction orders, $\lambda_1$ = 8.0135 μm and $\lambda_2$ = 7.989 μm, respectively. **(d)** Real part of scattered electric field for $E_F$ = 0.2 eV (left) and 0.6 eV (right), normalized by the electric field amplitude of the incident plane wave. The angles of wavefronts coincide with the diffraction angles of 80° (+1st) and -80° (-1st). The outlines of the structure are shown in white lines.

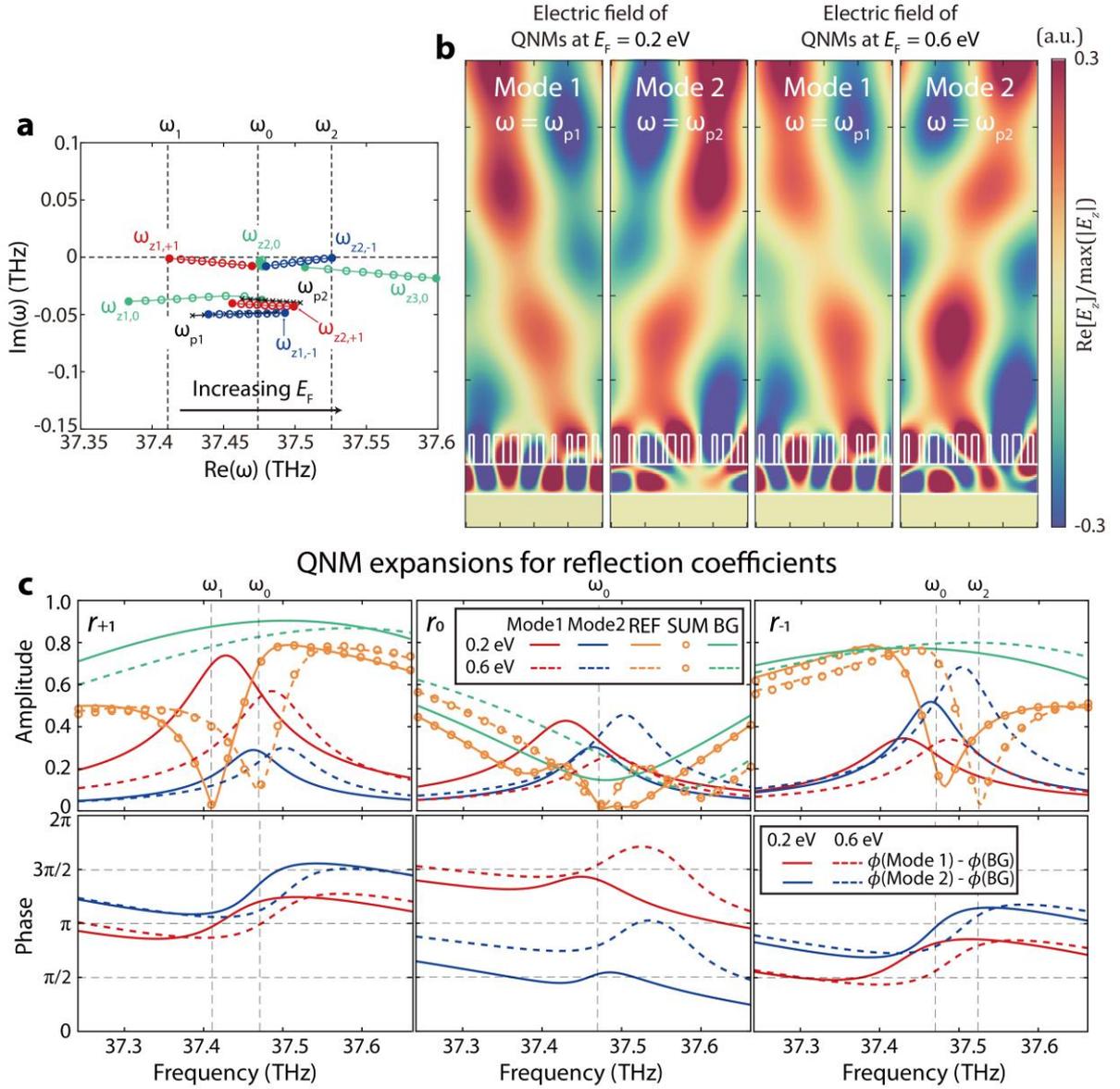

**Fig 3. Complex frequency analysis and QNM decomposition of the two-level beam switching metasurface. (a)** The locations of poles and zeros, depending on continuous graphene Fermi level range of $E_F$ = 0.2-0.6 eV, by a step of 0.05 eV. The poles are indicated by black 'X' marks. The zeros of +1st, 0th, -1st diffraction orders are indicated by red, green, and blue circles, respectively. The zeros at $E_F$ = 0.2, 0.6 eV are indicated as filled circles. The locations of frequencies $\omega_0$, $\omega_1$, and $\omega_2$ which correspond to wavelength $\lambda_0$, $\lambda_1$, and $\lambda_2$ are indicated as vertical dotted lines. **(b)** Real part of electric fields of QNMs at two switching states. Mode 1 corresponds to $\omega_{p1}$ and Mode 2 corresponds to $\omega_{p2}$. **(c)** QNM expansions of reflection (diffraction) coefficients. The amplitude of each contributing reflection coefficients, and their relative phases with respect to background mode are drawn in upper row and lower row, respectively. (Upper row) For each diffraction order, the contribution from mode 1, mode 2, and background mode is drawn in red, blue, and green curves, respectively. Reference and summed quantities are drawn in yellow curves and circles. (Lower row) Relative phases of mode 1 and mode 2 with respect to background mode are drawn in red and blue curves. For both rows, solid and dotted curves represent the result at $E_F$ = 0.2 eV and $E_F$ = 0.6 eV.

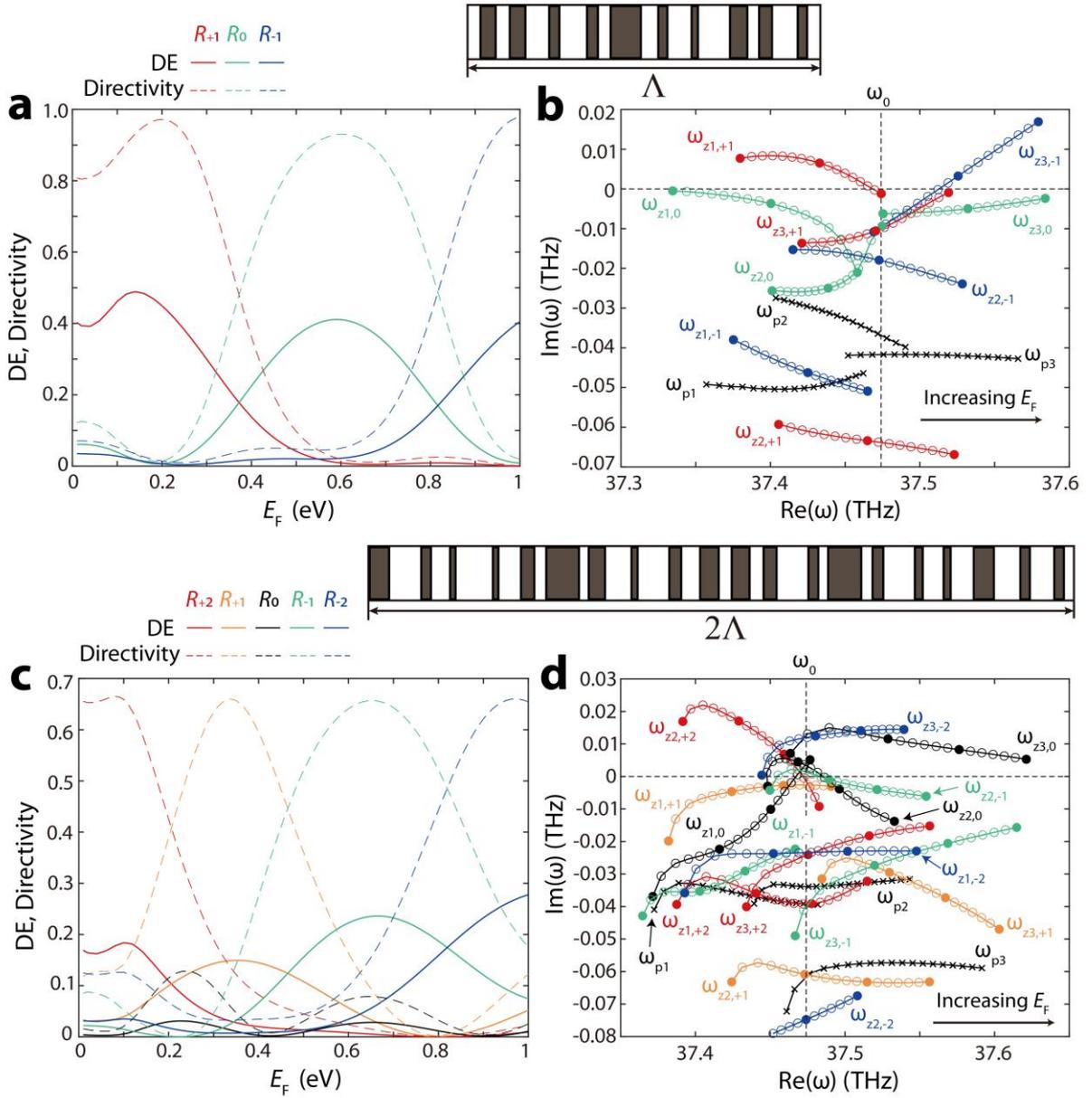

**Fig. 4. Multi-level beam switching metasurfaces. (a)** Diffraction efficiencies and directivities of the three-level beam switching metasurface, indicated by solid lines and dashed lines, respectively. Red, green, and blue lines correspond to the +1$^{st}$, 0$^{th}$, and -1$^{st}$ diffraction orders, respectively. **(b)** The locations of poles and zeros, dependent on the graphene Fermi levels of $E_F$ = 0.2-1.0 eV shown with steps of 0.05 eV. The poles are indicated by black 'X' marks. The zeros of +1$^{st}$, 0$^{th}$, -1$^{st}$ diffraction orders are indicated by red, green, and blue circles, respectively. The zeros at $E_F$ = 0.2, 0.6, 1.0 eV are indicated as filled circles. **(c)** Diffraction efficiencies and directivities of four-level beam switching metasurface, indicated by solid lines and dashed lines, respectively. Red, yellow, black, green, and blue lines correspond to the +2$^{nd}$, +1$^{st}$, 0$^{th}$, -1$^{st}$, and -2$^{nd}$ diffraction orders, respectively. **(d)** The locations of poles and zeros, dependent on the graphene Fermi level range of $E_F$ = 0.05-0.95 eV with steps of 0.05 eV. The poles are indicated by black 'X' marks. The zeros of the +2$^{nd}$, +1$^{st}$, 0$^{th}$, -1$^{st}$, -2$^{nd}$ diffraction orders are indicated by red, yellow, black, green, and blue circles, respectively. The zeros at $E_F$ = 0.05, 0.35, 0.65, 0.95 eV are indicated as filled circles. The optimized metagrating patterns for the three-level and the four-level switching metasurfaces are shown above **(a, b)** and **(c, d)**, respectively.

# Supplementary Information

# High-directivity multi-level beam switching with single-gate tunable metasurfaces based on graphene

Juho Park[1], Ju Young Kim[1], Sunghyun Nam[1], and Min Seok Jang[1*]

[1]School of Electrical Engineering, Korea Advanced Institute of Science and Technology, Daejeon 34141, Republic of Korea

*Corresponding author. Min Seok Jang: jang.minseok@kaist.ac.kr

# S1. Adjoint gradient formulation

Adjoint method is an efficient tool for calculating gradients with respect to design parameters. It is vastly used for many photonic inverse design applications. Adjoint method needs forward and adjoint (backward) simulations for calculating the gradient. For beam deflecting applications, the forward simulation is needed to evaluate the diffraction efficiency and the forward electric field $E_{\text{for}}$ inside the metagrating region, when the source is an incident plane wave. The adjoint simulation is needed to evaluate the adjoint electric field $E_{\text{adj}}$ when the source is a backward incident plane wave from the desired target direction. In this work, we utilize the shape-derivative version of the adjoint method to avoid additional binarization procedures, otherwise needed for grayscale permittivity gradient.

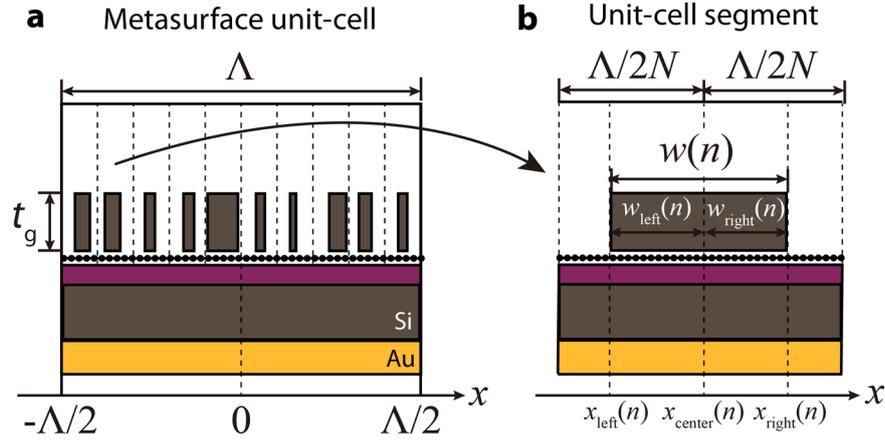

**Fig S1. (a) Metasurface unit-cell and (b) unit-cell segment.** The width of unit-cell segment drawn in **(b)** is exaggerated to illustrate parametrizations for shape-derivative.

To implement free-form optimizations using shape-derivatives, we first divide the metasurface unit-cell into $N = 10$ segments having equidistant widths, and place one Si pillar for each segment (**Fig. S1a**). And to make the Si pillar move freely inside each segment, we parametrize the left and right widths from the center of segment as follows (**Fig. S1b**):

$$w_{\text{left}}(n) = p_{\text{left}}(n)\frac{\Lambda}{2N}, \quad w_{\text{right}}(n) = p_{\text{right}}(n)\frac{\Lambda}{2N},$$

$$w(n) = w_{\text{left}}(n) + w_{\text{right}}(n), \quad n = 1, 2, \ldots, N,$$

where $p_{\text{left}}(n)$ and $p_{\text{right}}(n)$ represent normalized left and right widths of Si pillars. By setting the range of the normalized left and right widths between zero and unity, one can represent any Si pillar shape inside the segment. The *x*-coordinates of the center point and both edges of the Si pillar are represented as follows:

$$x_{\text{center}}(n) = \frac{\Lambda}{2N}(-N + 2n - 1),$$

$$x_{\text{left}}(n) = x_{\text{center}}(n) - w_{\text{left}}(n), \quad x_{\text{right}}(n) = x_{\text{center}}(n) + w_{\text{right}}(n)$$

Using the notations above, we get the adjoint gradient formulation as follows [1]:

$$\frac{\partial FoM}{\partial \boldsymbol{p}} = \frac{t_g \Lambda}{2N}\{[\epsilon_{Si} - \epsilon_{Air}][E_{for}^{\parallel}(\boldsymbol{x}(\boldsymbol{p})) \cdot E_{adj}^{\parallel}(\boldsymbol{x}(\boldsymbol{p})) + \frac{1}{\epsilon_{Si}\epsilon_{Air}} \cdot D_{for}^{\perp}(\boldsymbol{x}(\boldsymbol{p})) \cdot D_{adj}^{\perp}(\boldsymbol{x}(\boldsymbol{p}))]\}$$

where $E_{for}^{\parallel}, E_{adj}^{\parallel}$ represent forward and adjoint fields parallel to the boundary of the pillars, respectively, $D_{for}^{\perp}, D_{adj}^{\perp}$ represent forward and adjoint displacement fields perpendicular to the boundary of the pillars, respectively, and $\boldsymbol{x}(\boldsymbol{p})$ is a vector of the locations of the left and the right boundaries of the pillar, and $\boldsymbol{p}$ is a vector of normalized left and right widths of the Si pillars.

# S2. Metagrating layer and substrate thickness sweep analysis

In this section, we report the thickness sweep results (**Fig. S2**) for metagrating and substrate thicknesses, represented as $t_g$ and $t_s$ in **Fig. 1** in the main script. Sweep ranges for both thicknesses are $k_{Si}t_g = \pi/8 \sim 2\pi$ and $k_{Si}t_s = \pi/8 \sim 2\pi$, with steps of $\pi/8$. At each thickness pair ($k_{Si}t_g$, $k_{Si}t_s$), we conduct 5 optimizations and selected the optimized design having the maximum value of minimum directivities of all orders. We highlight the data points with over 90% minimum directivities, and draw average diffraction efficiencies for those data points. For both two- and three-level beam switching metasurfaces, designs with high directivities were achieved near ($k_{Si}t_g$, $k_{Si}t_s$) = ($3\pi/2$, $3\pi/2$). This concentration is more evident for the three-level case. For the four-level switching optimization, we could not achieve designs that displayed over 90% minimum directivities. Interestingly, we observed there is a "forbidden zone" where directivities are low for all three metasurfaces, indicated by red lines in **Fig. S2**. We suspect that the reason for this forbidden zone is possibly due to the Salisbury-screen effect that reduces direct reflection of normally incident waves. A Salisbury screen is one of the antireflection technologies used to eliminate incident waves [2]. It operates on the similar mechanism used for antireflection coatings, based on inducing a $\pi$ phase shift between the incident wave and the reflected wave, allowing for destructive interference. Here it is desirable to represent phase shifts due to the metagrating and the substrate into $\Delta\phi_{MG} = k_{Si}t_g$ and $\Delta\phi_{Sub} = k_{Si}t_s$. Considering that in our metasurface setup the round-trip phase shift between the incident and the outgoing wave from the metasurface is approximately $2k_{Si}(t_g + t_s) = 2(\Delta\phi_{MG} + \Delta\phi_{Sub})$ (ignoring the phase shift from the graphene and the Si$_3$N$_4$ spacer), the antireflection will occur at $\Delta\phi_{MG} + \Delta\phi_{Sub} = (2n + 1)\pi/2$, where $n$ is a non-negative integer. For zone #1 ($n$ = 0), a one-way phase shift due to the metasurface is below $\pi/2$, i.e. $\Delta\phi_{MG} + \Delta\phi_{Sub} \leq \pi/2$ and $\Delta\phi_{Sub} \leq \pi/2$. Here, the condition for antireflection is not met, and directivities of the optimized metasurfaces are low. If we increase $\Delta\phi_{MG}$ while fixing $\Delta\phi_{Sub}$, i.e. $\Delta\phi_{MG} + \Delta\phi_{Sub} \geq \pi/2$ and $\Delta\phi_{Sub} \leq \pi/2$, the condition is met and the primary objective is achieved, and the remainder of the freedom in the design parameter space seemed to be devoted to the beam shaping of the other diffraction channels. This phenomenon may be related to forming more suitable Bloch modes resident in the metagrating region, having a larger outcoupling to target diffraction channels, as reported in [3]. For zone #2 ($n$ = 1), the range of $\Delta\phi_{Sub}$ becomes $\pi/2 \leq \Delta\phi_{Sub} \leq 3\pi/2$, unable to satisfy the first phase matching condition ($n$

= 0). Therefore, the directivities remain low until the second phase matching condition $\Delta\phi_{MG} + \Delta\phi_{Sub} = 3\pi/2$ is satisfied. When $\Delta\phi_{MG}$ is increased to match the second phase matching condition and beyond, again for the same reason, directivities improve. Consequently, these periodic phase matching conditions form a sawtooth pattern enclosing Forbidden zones, shown in **Fig. S2**.

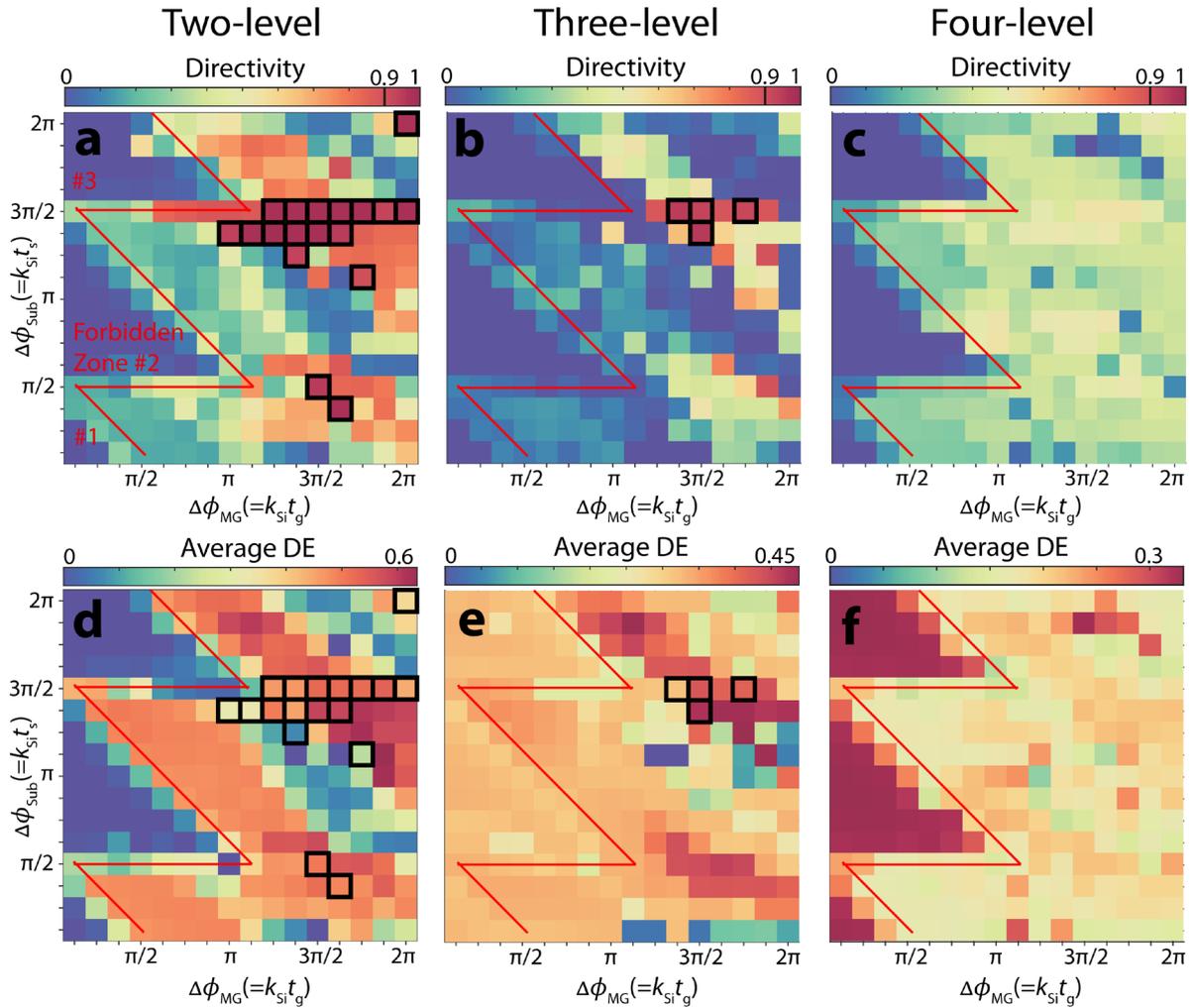

**Fig S2. Thickness sweep results for directivity and average diffraction efficiency for optimized (a, d) two-, (b, e) three-, and (c, f) four-level beam switching metasurfaces.** Data points enclosed with solid squares represent the optimized designs with over 90% directivities. The "forbidden zone" for low directivities is indicated by red lines.

# S3. Band structure analysis

In the main text, we discussed how the quasinormal modes can contribute to forming active beam switching. Here we present band structures of two- and three-level optimized metasurfaces along with the angle-resolved absorption spectra. Band structures are obtained with eigenfrequency analysis using FEM software. Absorption spectra are calculated with RETICOLO RCWA software [4]. For two-level case, there are two poles (eigenfrequencies) concentrated near the operating frequency $\omega_0$ (corresponding to $\lambda_0 = 8$ μm), rapidly diverging along increasing incident angles. Similar pole behaviors are observed in the three-level case as well. These pole concentrations and angle-sensitive diverging imply that the quasinormal mode resonances are very sophisticated in terms of arrangement with the help of numerical optimization, and imply that such could not be easily achieved with conventional design procedures. Also, the blueshift of poles along increasing Fermi levels is observed, which is expected from the linear resonance frequency shift discussed in **Section 2.2**.

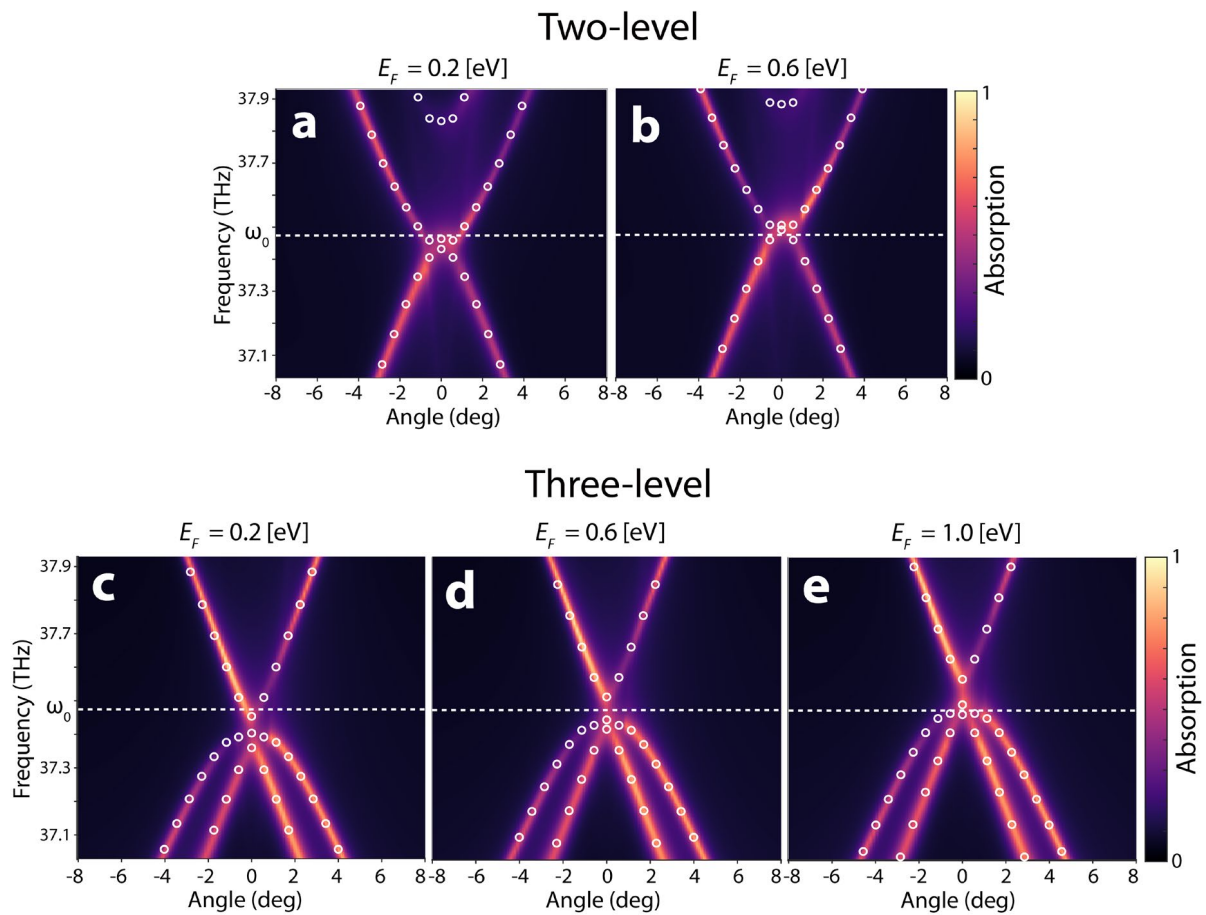

**Fig S3. Band diagrams showing angle-resolved absorption spectra of (a, b) two- and (c, d, e) three-level switching metasurfaces.** White circles indicate eigenfrequencies obtained from eigenfrequency

analysis.

# S4. QNM expansion of three-level switching metasurface

Using the same procedure for the two-level case, QNM expansion was applied to the three-level metasurface. The main difference between the two-level and the three-level is that an additional QNM (pole) was considered for the expansion. Unlike the two-level case, we could not find a clear relationship between the background and each QNM, which was previously evident based on relative phases. Instead, three QNMs interact with each other to form high directivities at each Fermi level. The numbering of modes are consistent with **Fig. 4** in the main text. Note that the over-unity reflection amplitude of mode 2 for the $0^{th}$ reflection channel at Fermi level 1.0 eV is a result of a mathematical byproduct of the Riesz Projection, which might result in an unphysical quantity [5-7].

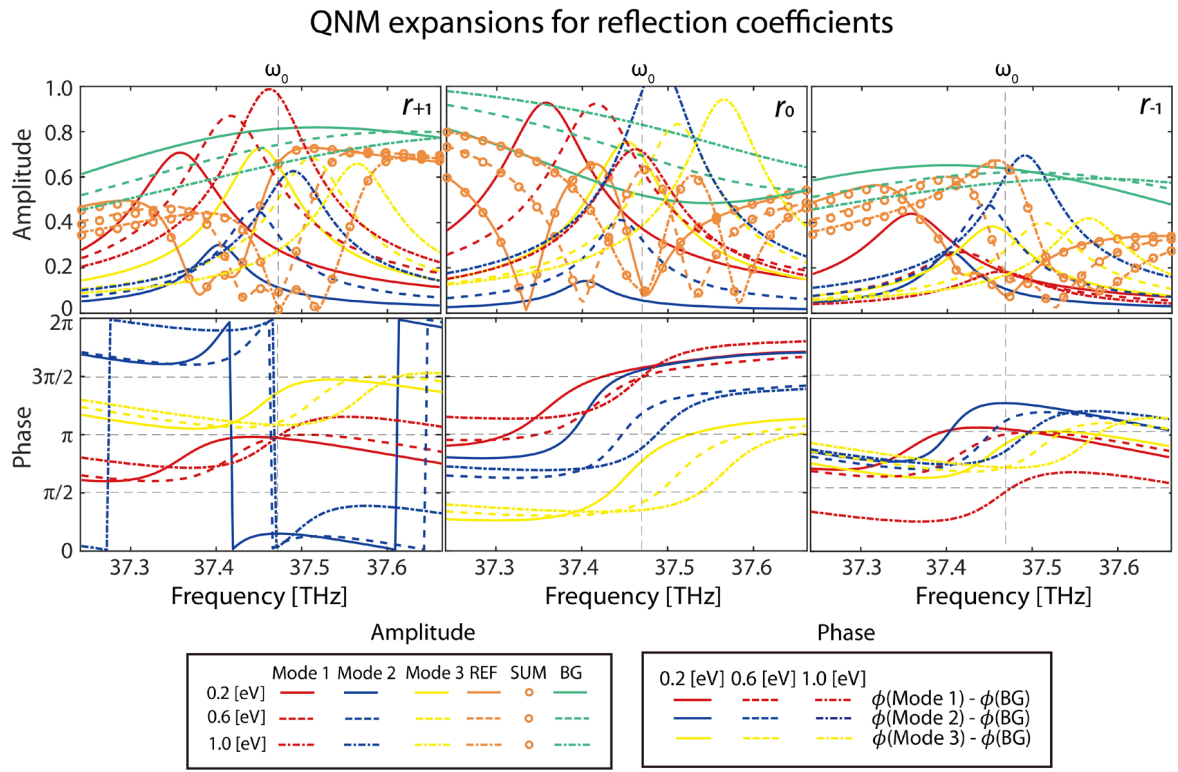

**Fig S4. QNM expansion results for three-level beam switching metasurface.**

# S5. Inverse-designed metagrating patterns of the metasurfaces

We tabulate the left edge locations and widths of the Si metagrating using the variables illustrated in **Fig. S1b**. Note that for the two- and three-level switching metasurfaces, the unit-cell period is $\lambda_0/\sin(80°) = 8.123$ μm and $N = 10$. For the four-level switching metasurface, both the unit-cell period and the number of pillars are doubled.

**Table S1. Left edge locations and widths of Si pillars for optimized metagrating patterns.**

| $n$ | Two-level | | Three-level | | Four-level | |
|---|---|---|---|---|---|---|
| | $x_{\text{left}}(n)$ (μm) | $w(n)$ (μm) | $x_{\text{left}}(n)$ (μm) | $w(n)$ (μm) | $x_{\text{left}}(n)$ (μm) | $w(n)$ (μm) |
| 1 | -3.8033 | 0.2557 | -3.7881 | 0.3972 | -8.1188 | 0.4962 |
| 2 | -2.9269 | 0.2785 | -3.1131 | 0.4077 | -6.9303 | 0.2710 |
| 3 | -2.3566 | 0.5475 | -2.2058 | 0.2831 | -6.2680 | 0.1754 |
| 4 | -1.3688 | 0.4243 | -1.3352 | 0.299 | -5.2802 | 0.1754 |
| 5 | -0.7019 | 0.4891 | -0.7927 | 0.7481 | -4.6338 | 0.3562 |
| 6 | 0.1662 | 0.3951 | 0.304 | 0.2593 | -4.0526 | 0.8032 |
| 7 | 1.0881 | 0.1974 | 1.0739 | 0.1946 | -3.0739 | 0.4250 |
| 8 | 1.9644 | 0.275 | 1.9579 | 0.4477 | -2.0892 | 0.1892 |
| 9 | 2.5894 | 0.4931 | 2.629 | 0.3458 | -1.2185 | 0.3229 |
| 10 | 3.4656 | 0.3147 | 3.5176 | 0.2649 | -0.5091 | 0.4774 |
| 11 | | | | | 0.2194 | 0.4583 |
| 12 | | | | | 0.9522 | 0.3486 |
| 13 | | N/A | | | 1.9843 | 0.2771 |
| 14 | | | | | 2.4370 | 0.8123 |
| 15 | | | | | 3.4621 | 0.2955 |
| 16 | | | | | 4.4362 | 0.2286 |
| 17 | | | | | 5.0984 | 0.2083 |
| 18 | | | | | 5.7920 | 0.5132 |
| 19 | | | | | 6.8638 | 0.2691 |
| 20 | | | | | 7.6457 | 0.2697 |

## S6. Graphene permittivity dependence on Fermi level

In this section, we discuss the dependency of the graphene permittivity on the Fermi level $E_F$. To obtain the permittivity, we first calculate the conductivity of graphene following the derivation based on [8]. Then, the conductivity was converted to isotropic permittivity. The electron mobility values discussed are $\mu_s$ = 500, 1000, 1500, 2000 cm$^2$/V·s. In **Fig. S5**, for $E_F$ < $\hbar\omega/2$ ~ 0.0775eV, the high values of Im{$\varepsilon_{graphene}$} are due to interband transition losses. As $E_F$ increases, more intraband transition takes place, shown in a bulge and a dip in Re{$\varepsilon_{graphene}$} and Im{$\varepsilon_{graphene}$}, respectively. After some point around $E_F$ = 0.2 eV, the intraband transition seems to settle in, and Re{$\varepsilon_{graphene}$} exhibits a linear relationship with $E_F$. Meanwhile, Im{$\varepsilon_{graphene}$} converges to a constant value as $E_F$ increases, but these values decrease with increasing $\mu_s$. The linearity of Re{$\varepsilon_{graphene}$} and the constant values of Im{$\varepsilon_{graphene}$} along $E_F$ lead to linear resonance frequency shifts of eigenmodes with preserved linewidths within our $E_F$ range of interest, explaining the reason for the preserved lineshapes in **Fig. 2c**.

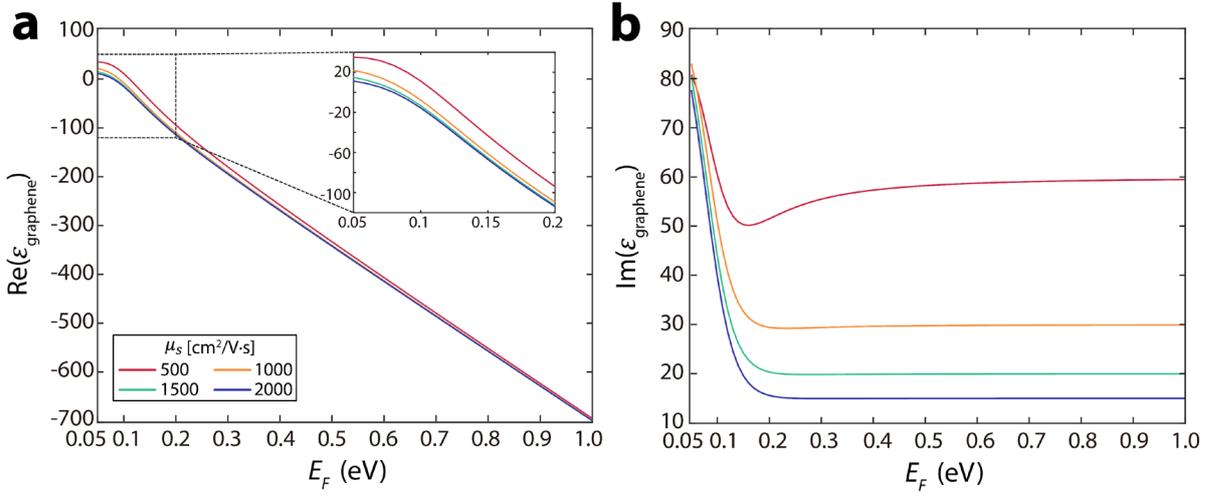

**Fig S5. Graphene permittivity dependence on Fermi levels for $\mu_s$ = 500, 1000, 1500, 2000 cm$^2$/V·s.**

# S7. Riesz Projection method

One way to analyze the relationship between the reflectance/transmittance spectrums with multiple resonances is to utilize temporal coupled-mode theory [9,10] (TCMT). Although TCMT is a general, well-founded multi-port resonator model, we found applying TCMT in our 3-port (=the number of diffraction channels) metasurface system to be difficult, since the 3-by-3 coupling coefficient matrix needed to establish TCMT resulted in too many undetermined variables [11]. For our work, we employ RP methods that yield a clearer analysis for our optical system.

A brief explanation of the Riesz Projection methods is stated hereafter. We define $q(\omega)$ as the analytical continuation of the physical observable of interest. Deploying the Cauchy's residue theorem on $q(\omega)$ and manipulating the contour so that the resonance poles of concern are enclosed by their respective contours, we get the following expansion [12,13]:

$$q(\omega) = \frac{1}{2\pi i} \oint_C \frac{q(\xi)}{\xi - \omega} d\xi = \sum_{m=1}^{M} q_m(r, \omega) + q_{nr}(r, \omega), \text{where}$$

$$q_m(r, \omega) = \frac{-a_m}{\omega_m^{pole} - \omega}, a_m = \frac{1}{2\pi i} \oint_{C_m} q(\xi) d\xi, q_{bg}(\omega) = \frac{1}{2\pi i} \oint_{C_{nr}} \frac{q(\xi)}{\xi - \omega} d\xi \quad (2)$$

Here, the contour $C$ contains inside the frequency of interest $\omega$, but not the pertinent poles. The $q_m(r,\omega)$ embodies the RP onto the eigenspace of each eigenvalue/pole, representing the contribution to the spectrum from each mode. The residues $a_m$ are evaluated along the contours $C_m$, which involve only one corresponding pole inside. The $q_{bg}(\omega)$ gives the background contribution including the non-resonant effects and the influence of the poles outside the main outer contour $C_{nr}$.